\documentclass[]{elsarticle}

\usepackage{lineno,hyperref}
\modulolinenumbers[5]
\usepackage{amsmath}
\usepackage{color}
\usepackage{lscape}
\usepackage{graphicx}
\usepackage{pgfplots}
\pgfplotsset{compat=1.7}

\usepackage[acronym, style=super, nonumberlist]{glossaries}
\newacronym{eeg}{EEG}{electroencephalogram}
\newacronym{ml}{ML}{machine learning}
\newacronym{cv}{CV}{cross validation}

\journal{Journal of \LaTeX\ Templates}

\begin{document}

\title{Nonlinear and Machine Learning Analyses on High-Density EEG data of Math Experts and Novices}

\author{Hanna Poikonen, Tomasz Zaluska, Xiaying Wang, Michele Magno, Manu Kapur\\
Co-first-authors: H. Poikonen, T. Zaluska}

\maketitle

\begin{abstract}
Current trend in neurosciences is to use naturalistic stimuli, such as cinema, class-room biology or video gaming, aiming to understand the brain functions during ecologically valid conditions. Naturalistic stimuli recruit complex and overlapping cognitive, emotional and sensory brain processes. Brain oscillations form underlying mechanisms for such processes, and further, these processes can be modified by expertise. Human cortical oscillations are often analyzed with linear methods despite brain as a biological system is highly nonlinear. This study applies a relatively robust nonlinear method, Higuchi fractal dimension (HFD), to classify cortical oscillations of math experts and novices when they solve long and complex math demonstrations in an EEG laboratory. Brain imaging data, which is collected over a long time span during naturalistic stimuli, enables the application of data-driven analyses. Therefore, we also explore the neural signature of math expertise with machine learning algorithms. There is a need for novel methodologies in analyzing naturalistic data because formulation of theories of the brain functions in the real world based on reductionist and simplified study designs is both challenging and questionable. Data-driven intelligent approaches may be helpful in developing and testing new theories on complex brain functions. Our results clarify the different neural signature, analyzed by HFD, of math experts and novices during complex math and suggest machine learning as a promising data-driven approach to understand the brain processes in expertise and mathematical cognition.

\end{abstract}



\begin{keyword}
\texttt Math \sep Expertise \sep EEG \sep  Higuchi Fractal Dimension \sep Machine learning \sep Naturalistic stimulus
\end{keyword}


\section{Introduction}

For decades, brain research has aimed to decode the brain states in development, illness and health to understand normal and abnormal brain functions. Current trend in neurosciences is to use naturalistic stimuli which aims to understand the brain functions in the real world during which sensory, cognitive, emotional and motor brain processes overlap(Sonkusare et al., 2019 \cite{SONKUSARE2019699}; Cantlon, 2020 \cite{Cantlon}; Nastase et al., 2020 \cite{Nastase2020KeepIR}; Zhang et al., 2021 \cite{Zhang2021NaturalisticSA}). Naturalistic stimuli mean complex, dynamic and diverse stimuli which create a more ecologically relevant condition for brain research in comparison to the traditionally used reductionist stimuli (Cantlon, 2020\cite{Cantlon} ; Zhang et al., 2021 \cite{Zhang2021NaturalisticSA}). Examples of naturalistic stimuli are cinema, classroom biology, video gaming, complex math or listening to a live orchestra (Hasson et al., 2004 \cite{Hasson2004IntersubjectSO} ; Dikker et al., 2017 \cite{Dikker2017BraintoBrainST}; Bavelier and Green \cite{Bavelier2019EnhancingAC}, 2019; Chabin et al., 2022 \cite{Chabin2022}; Poikonen et al., 2022 \cite{poikonnen2022}).

Continuous brain imaging data, which is collected over a long time span during naturalistic stimuli, enables the application of data-driven analyses (Cantlon, 2020 \cite{Cantlon}; Zhang et al., 2021 \cite{Zhang2021NaturalisticSA}). Machine learning (ML) analyses may assist in generating new hypotheses about the underlying task-relevant brain processes, especially in the naturalistic context. In such contexts, several low and high-level overlapping brain processes occur simultaneously (Nastase et al., 2020 \cite{Nastase2020KeepIR}). Due to the overlapping nature of several brain processes, extension of the neuroscientific theories formulated based on reductionist and simplified study designs is both challenging and questionable (Cantlon, 2020 \cite{Cantlon}). Novel methodologies in analyzing naturalistic data are required and data-driven intelligent approaches form a good candidate for developing and testing new theories on the brain functions in the real world (Nastase et al., 2020 \cite{Nastase2020KeepIR}).

Recent developments in ML are already applied in healthcare and extend to several fields: spike detection in epilepsy, dementia prediction, and mental health and sleep stage classification (Singh et al., 2022 \cite{singh}). 
These data-driven methods aim to transform healthcare delivery and to change the trajectory of brain health by addressing brain care earlier in the lifespan (Singh et al., 2022 \cite{singh}). For example, recent advances utilizing ML, specifically techniques with Brain-Computer Interfaces (BCI), help stroke patients either restore neurologic pathways or communicate with an electronic prosthetic (Cervera et al., 2018 \cite{Cervera2017BrainComputerIF}; Baniqued et al., 2021 \cite{Baniqued2019BraincomputerIR}). On the other hand, ML may help in diagnostics of conditions like stroke or Alzheimer through the detection of disease-specific EEG biomarkers (Karthik et al., 2020 \cite{KARTHIK2020105728}; Meghdadi et al., 2021 \cite{Meghdadi2021}).

In addition to the applications for prediction and diagnostics in healthcare, ML for brain imaging has application possibilities in the contexts of learning and education (Bavelier and Green, 2019  \cite{Bavelier2019EnhancingAC}; Cantlon, 2020 \cite{Cantlon}). For decades, scientists have studied the brain processes during cognitive tasks, like mathematics or language. These studies have brought valuable knowledge on the domain-general brain functions of working memory, attention, and solving strategies (e.g. De Smedt et al., 2009 \cite{smedt}; Kulasingham et al., 2021 \cite{Kulasingham}; Wang et al., 2020 \cite{Wang2020}) and domain-specific brain functions on numeric and verbal processing (e.g. Amalric and Dehaene 2016 \cite{Amalric2016}, 2019 \cite{AMALRIC2019}). Some studies have focused to understand healthy development and expertise (Jeon et al., 2019 \cite{Jeon2019}; Zhang et al., 2015 \cite{Zhang2015}), whereas others bring insights on disrupted development and learning deficits (Klados et al. 2017 \cite{KLADOS201724}; Rubinsten, 2015 \cite{Rubinsten}). Neuroscientific studies made in learning sciences have not yet utilized ML in the data analysis. However, ML has potential to be used in data-driven hypothesis formation of the brain functions underlying expertise development or learning deficits, and for real-time adaptive feedback in learning and focused attention (Kefalis et al., 2020 \cite{Kefalis2020TheEO}; Hunkin et al., 2021 \cite{Hunkin2020EEGND}). 

Previous studies show differences in the brain functions of math experts and novices during short and simple math tasks (e.g. Grabner et al., 2007 \cite{Grabner2007}). Such differences are associated with brain functions modified through expertise, such as rote learning and strategy selection for solving the tasks at hand (Grabner and De Smedt, 2012 \cite{Grabner2012}; Hinault and Lemaire, 2016 \cite{Hinault2016}). However, a few second simple math tasks, which are used traditionally as stimuli in studies on math expertise, seldomly create enough of continuous brain imaging data for which to successfully apply the ML methods.  

Despite some ML algorithms are designed to evaluate raw EEG data (da Silva Lourenco et al., 2021 \cite{Silva2021}), several studies which focus on the comparison of brain states have preprocessed the data before ML classifications. The brain, as many biological systems, behaves in a nonlinear manner. Nonlinear behavior of biological systems is characterized by a high degree of variability in the time domain (nonstationarity) and randomness that could be attributed to the interaction of internal and external factors influencing the organism (Glass, 2001 \cite{Glass2001}; Eke et al., 2002 \cite{Eke2002}). Engagement with complex math recruits several cognitive brain processes which overlap with sensory and emotional processes (Suarez-Pellicioni et al., 2016 \cite{Suarez2015}; Wang et al., 2015 \cite{Wang2015}). Therefore, the EEG data collected during such cognitively challenging task is likely highly complex, and therefore, a potentially optimal way to process such data includes an analysis which is suitable for nonlinear systems.  

Cognitively challenging tasks create a brain state which is clearly different from those of relaxed states (Finn, 2021 \cite{Finn2021}). Fractal dimension is a highly sensitive measure in the detection of hidden information contained in physiological time series (Klonowski \cite{Klonowski2002}, 2002; Raghavendra and Dutt, 2010 \cite{RAGHAVENDRA2010}) and is shown to vary depending on the brain state. An often-used nonlinear measure for signal analysis is Higuchi’s fractal dimension (HFD) which is a measure of signal complexity in the time domain (Higuchi, 1988 \cite{HIGUCHI1988277}; Spasic et al., 2008 \cite{Spasic2008}). Previous studies utilizing such methods classified successfully different sleep stages and detected the difference in the brain state during drowsiness and wakefulness (Inoye et al., 1994 \cite{Inouye1994ChangesIT}; Šušmáková and Krakovská, 2008 \cite{Susmakova2008}). HFD showed the most robust results and seems to be superior to other FD methods for EEG signals (Solhjoo and Nasrabadi, 2005 \cite{Solhjoo2005}; Ahmadlou et al., 2012 \cite{AHMADLOU2012206}). 

This study investigated the neural signature of math expertise with a relatively robust nonlinear analysis, HFD, and explored a new paradigm by applying ML to EEG data collected from math experts and novices when they engaged with long and complex math demonstrations. Such math demonstrations with a duration up to one minute form a part of the current trend in investigating the brain with naturalistic stimuli. Our aim was to describe the EEG data during advanced mathematical cognition with a nonlinear method and evaluate whether the neural signature of math experts and novices differ in a way which is detectable with artificial intelligence. We hypothesized that the experts’ and novices’ brain functions during long math tasks differ in signal complexity detectable with HFD, which further, can be classified by a ML model. 

\section{Materials and methods}

\subsection{Participants}
Thirty-four math experts (bachelor and master students in math or math-related disciplines, like physics or engineering) and thirty-five math novices (no university-level math studies) participated in the experiment. However, eleven participants from the group of math experts and twelve participants from the novice group were discarded from the data analysis because their EEG data was too noisy, or some of the relevant data was missing due to malfunctioning EEG amplifier. Therefore, in the group of math experts, there were 22 participants (5 female and 17 male), and in the novice group, 22 participants (7 female and 15 male). The background of the participants was screened by a math questionnaire.

The age of the participants ranged from 19 to 24 years (mean 21.0 years) among math experts and from 19 to 35 years (mean 23.8 years) among novices. All participants in both groups were right-handed. No participants reported hearing loss nor history of neurological illnesses. The experiment protocol was conducted in accordance with the Declaration of Helsinki and approved by the Executive Board of ETH Zurich after a review by the ETH Zurich Ethics Commission. All participants provided written informed consent.

\subsection{Task design}
Participants watched 16 math demonstrations. After each demonstration they were asked three self-evaluation reflections to which they answered by pressing a button in a 4-button response box. Each set of trials consisted of four excerpts of the same presentation style (symbolic or geometric), and these sets were presented in a pseudo random order via a monitor. The pseudo randomization defined the presentation order (symbolic first or geometric first). However, each participant saw the same four math demonstrations presented in both symbolic and geometric form before seeing them in the other form.  

Each math demonstration consisted of several slides, from 4 up to 12 slides (6.9 slides on average) depending on the complexity of each demonstration. The total duration of math demonstrations varied from 13 seconds to 68 seconds (33.1 seconds on average). The timing of each slide was the same for all the participants. The duration of each slide was defined according to an online screening in which 25 math experts and 25 math novices watched the math demonstrations slides and auto-regulated the following slide with a button press. The participants who attended the online screening did not attend the actual EEG experiment. The duration of each slide in the EEG experiment was the average time the participants spent on each slide during the online screening. In the online screening, there was no statistically significant difference between experts and novices in the duration of time spent on each slide.

\subsection{Data acquisition}
The stimuli were presented to the participants with the MATLAB via PsychToolbox. The experimenter launched the playback of the presentation program after which participant could navigate to the math demonstrations by a button press once they had read the instruction slides on the screen. The total length of the experiment material was approximately 15 minutes.

The data were recorded using Ant Neuro eego mylab electrode caps with active 128 EEG channels \footnote{\url{https://www.ant-neuro.com/products/eego_mylab}}. 

Four external electrodes placed below, above and on the left side of the left eye and on the right side of the right eye. The offsets of the active electrodes were kept below 30 mv at the beginning of the measurement, and the data were collected with a sampling rate of 2048 Hz. A timestamp (trigger) was marked into to EEG data at the beginning of each slide of the math presentations. The triggers were sent wirelessly via Lab Streaming Layer\footnote{\url{https://github.com/sccn/labstreaminglayer}}.

\subsection{Data pre-processing}
The EEG data of all the participants were first preprocessed with EEGLAB (version 2019.1; Delorme \& Makeig, 2004 \cite{Delorme2004}). The reference was set as the average of all the EEG electrodes. The data were high-pass filtered at 0.5 Hz and low-pass filtered at 40 Hz. Finite impulse response (FIR) filtering, based on the firls (least square fitting of FIR coefficients) MATLAB function, was used as a filter for all the data. Then, the data were treated with independent component analysis (ICA) decomposition with the runica algorithm of EEGLAB (Delorme and Makeig, 2004 \cite{Delorme2004}) to detect and remove artefacts related to eye movements and blinks. ICA decomposition gives as many spatial signal source components as there are channels in the EEG data. Typically, one to four ICA components related to the eye artefacts were removed. Noisy EEG data channels for some participants were interpolated. 

\subsection{Feature extraction}

\subsubsection{Higuchi Fractial Dimension (HFD)}
The EEG time-series has a duration between 10-20 minutes, resulting in a large data size per sample. Hence, feature extraction is necessary to capture relevant information. The extracted feature are then used to draw conclusions regarding the relevance of each brain area for mathematical calculations. For this purpose the fractal dimension (FD) \cite{Burns2015} for each sample is calculated and is used to measure the complexity of the signal. A simple pattern that is repeating continuously can become a very complex series which is the basis for the fractal constructs. A fractal is a shape that retains its structural detail despite scaling and is the reason why complex objects can be described with the help of fractal dimension. One variant of FD, the Higuchi's fractal dimension, \cite{HIGUCHI1988277} has its roots in chaos theory and has been successfully applied as a complexity in various domains of signal processing. It has been shown to be a good numerical solution to nonlinear signals \cite{KESIC201655}.
The speed, accuracy, and cost of applying the HFD method for research and medical diagnosis make it stand out from the widely used linear methods \cite{kawe}. Among the different FD algorithms Higuchi’s method \cite{KESIC201655} demonstrates to be a more accurate option for EEG signals, since it is accurate for stationary and non-stationary signals. 

Say $\mathbf{X}$ is an EEG signal of length $T$ is the length of a time window on which we calculate a HFD value. Following \cite{HIGUCHI1988277}, we calculate HFD as:
A new signal $x_m^{k}$ is constructed from $\mathbf{X}$, with window size N where m = (1, 2, ..., k) denotes the starting point and k = (1, 2, ..., $k_{max}$ ) the interval size:
\begin{align}
    x_m^{k} = \left \{ x(m), x(m+k), x(m+2k), ... , x(m+ \Bigl \lfloor  \frac{ N - m} { k } \Bigr \rfloor) \right\}
    \label{equ:HFD}
\end{align}
$L_m(k)$ describes the length of the curve of $x_m^{k}$ for every k given m:
\begin{align}
    L_m(k) = \frac{\sum_{i=1}x(m+ik)-x(m+(i-1)k)(N-1)}{\Bigl \lfloor  \frac{ N - m} { k } \Bigr \rfloor k}
\end{align}
where $\frac{N-1}{\Bigl \lfloor  \frac{ N - m} { k } \Bigr \rfloor}$ is the normalization factor. 
Length $L(k)$ is defined by the average of the k lengths:

\begin{align}
    L(k) = \frac{1}{k}\sum_{m=1}^{k}L_m(k)
    \label{equ:HFD_sum}
\end{align}

HFD is the slope of the best fitted curve between all the data points of time-series X for a given time window N for for k = (1, 2, ..., $k_{max}$ ) between log(1/k) and log L(k):

\begin{align}
    HFD(N,k_{max}): \text{best fit of}\  \left \{(log(\frac{1}{k}), log(k) \right\}
    \label{equ:HFD2}
\end{align}

It is possible to calculate HFD for the whole signal $(T=N)$. However, this is not recommended if the signal is nonstationary. In such cases the HFD value does not represent the true measure, and division into windows (or segments) is advised.
In \cite{accardo}, Accardo and colleagues have shown on synthetic fractal signals that Higuchi’s algorithm is more efficient,
faster, more accurate and able to estimate fractal dimension for short segments, compared to Maragos and Sun’s algorithm proposed in \cite{maragos}.

\subsubsection{Hyperparameter tuning}
An important hyperparameter that requires finetuning is $k_{max}$. There is no agreed methodology to optimize this parameter \cite{wanliss}. As per equation \ref{equ:HFD_sum}, HFD is summed up to $k_{max}$, therefore increasing $k_{max}$ will lead to an increase in HFD. A poor choice of $k_{max}$ will result in uninformative HFD, thus, it has to be carefully tuned.  

We propose the following methodology to identify the best value for $k_{max}$:
\begin{enumerate}
    \item We compute the HFD values as per equation \ref{equ:HFD2} for a wide range of $k_{max}$ values, i.e., $k_{max} \in {2, 5, 20, 100, 150, 200, 400}$ over all subjects and presentations.
    \item We identify the $k_{max}$ at which the difference (equation \ref{equ:diff}) between HFD values of significant and non-significant channels is maximized. Significance/non-significance is assessed by taking the maximum/minimum HFD value across all electrodes for a subject. Here, the minimum value is understood as the baseline fractal dimension and is therefore subtracted from the maximum value, which is the complexity of the relevant channels. We base this requirement on the assumption that certain EEG regions are more relevant than others for the mathematical tasks. Hence, there will be a difference in HFD values and we want to select the $k_{max}$ that maximizes this difference.
    \item The $k_{max}$ value that satisfies requirement 2) and 3) is chosen to compute the HFD values for further analyses and for the machine learning classification.
\end{enumerate}

\subsubsection{HFD features analyses}
Estimating HFD values for each channel of each participant allows to investigate which brain areas are most active while performing mathematical tasks. Since HFD values have no physical interpretation, a relative comparison between two different groups is performed. 

First, a comparison between experts and novices is investigated, by taking the average of all HFD values of the expert group and the novice group and subtracting them from each other: 
\begin{align}
    \Delta HFD_{ch_i} = \overline{\overline{HFD_{expert_j, pres_k}}}_{ ch_i} - \overline{\overline{HFD_{novice_j, pres_k}}}_{ch_i},
    \label{equ:diff}
\end{align}
\\
where $j\in$\{1,...11\} is the index of experts and novices, respectively, $k\in$\{1,...16\} is the index of presentations and $i\in$\{1,...129\} is the index of EEG channels.

A one-sided t-test is calculated, testing whether there is a significant difference between the two groups. A visual heatmap of the difference between experts and novices based on equation \ref{equ:diff} is mapped onto the head for better qualitative interpretation.

Subsequently a more fine grained analysis is performed by comparing the difference between expert and novice for algebraic and geometric separately:
\begin{align}
     \Delta_{AG}HFD_{ch_i} = \overline{\overline{HFD_{expert_j, pres_{k_A}}}}_ {ch_i} - \overline{\overline{HFD_{expert_j, pres_{k_G}}}}_{ch_i},
     \label{equ:5}
\end{align}
\\
where $k_A$ and $k_G$ $\in$\{1,...8\} is the index of the algebraic and geometric presentations respectively. 

\subsubsection{Machine learning classification}
We posit the question if a prediction can be made whether a new subject is a novice or an expert based on EEG recordings while performing mathematical tasks. We frame this problem as a two-class classification task. To understand and interpret the outcome of the machine learning classifiers, care needs to be taken while generating the classification dataset and splitting it into training and testing sets. 

We first define the classification-dataset as a collection of subject-presentation pairs (e.g. Expert1-Presentation1A etc.). Together with the 16 presentations, the full dataset include 704 samples, i.e., subject-presentation pairs. Subsequently, we calculate either a unique HFD value per EEG channel, meaning that each sample consists of 124 HFD features, or divide the EEG signals of total length T into non-overlapping windows of length N and calculate a HFD value for each window leading to (T/N)*124 HFD features. To be noted that the channels "VEOGL", "HEOGL", "HEOGR", "VEOGU", "HEART" are discarded, since they do not record brain signals but  eye movement and cardiac activity. 

\begin{figure}[!h]
\begin{center}
   \includegraphics[width=1.0\linewidth]{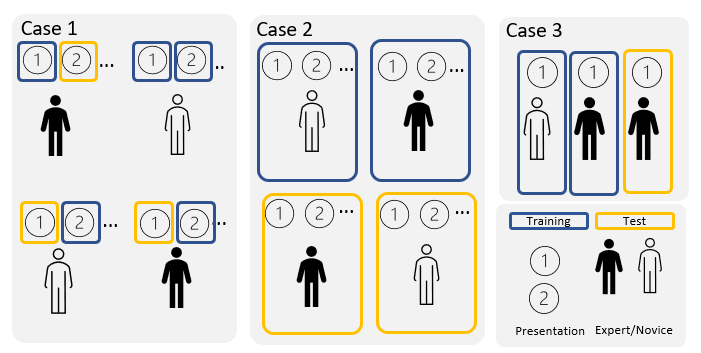}
\end{center}
   \caption{Classification-dataset split illustration. Case 1: Subject-presentation pairs split, Case 2: Subject-specific split, Case 3: Presentation-specific split.}
\label{fig:split}
\end{figure}

Since this work is the first in the literature to attempt an automatic classification of mathematical cognitive behavior, we propose three different cases of dataset splitting, illustrated in Figure~\ref{fig:split}:
\begin{enumerate}
    \item Subject-presentation pairs: We randomly split all 704 samples without considering whether a sample is coming from different subjects. This means that the samples from the same subject can either be entirely in the training set or in the validation set, or partially in the training and in the validation set.
    \item Subject-specific: We split the dataset on the level of subjects, meaning that all subject-presentation pairs of the same subject are either in the training or validation set.
    \item Presentation-specific: We deal with each presentation as a separate machine learning task. In other words, we divide the full dataset into sub-datasets, each of which consists in a single presentation, and perform the training and testing procedure on each of the sub-datasets.
\end{enumerate}

With case 1, we verify if the \gls{ml} classifier is able to discern between the 22 experts and novices present in the dataset based on a single mathematical presentation. With case 2, we validate the \gls{ml} classifier on new subjects of which data it has never seen before. The former is a relatively easier classification task, but necessary as a first proof-of-concept, whereas the latter tackles the most challenging problem of inter-subject variability common to all biomedical data.
%
%
With case 3, we  analyze whether a prediction can be made based on samples coming from a single presentation. By training a separate classifier for each presentation, we can compare the classification accuracy among the presentations and draw insights about which mathematical presentation is more suitable for discerning between math novices and experts. 

For cases 1 and 2 we calculate a single HFD value per EEG channel throughout the whole duration of the presentations. This choice is motivated by the fact that all presentations, of different recording lengths, belong to the same dataset on which a machine learning classifier is trained on and, in general, the classifiers require a fixed numbers of features.
This is no longer an issue for case 3, because each sub-dataset consists of data from a single presentation of fixed length. Hence, we can increase the granularity and use a non-overlapping moving window of length N to calculate the HFD value in equation \ref{equ:HFD2} for each window. More precisely, a HFD value is calculated every N seconds of the duration of the presentation $HFD_{1:N}, ... , HFD_{t:t+N}$ with $t$ being time steps. This allows to analyze the temporal evolution of the presentation and draw conclusions regarding the classification differences. We test several values of N, i.e., {5, 8, 11 seconds.}

Once the datasets are prepared, we proceed with classifiers training using the scikit-learn Python package. We investigate several \gls{ml} algorithms including Nearest Neigbours, Linear SVM, Decision Tree and Adaboost. We first optimize the classifiers by tuning the hyperparameters under case 1, i.e., subject-presentation level. Once the optimal parameters are found, we keep them for case 2 and 3.
\begin{table}[!ht]
    \centering
    \begin{tabular}[width=1.0\linewidth]{|l|l|}
    \hline
        \textbf{Algorithm} & \textbf{Parameters} \\ \hline
        Nearest Neighbors & Number neighbors: 3, 5, 7, 9, 11 \\ \hline
        Linear SVM & Kernel: linear, C: 0.025, 0.5, 0.75 \\ \hline
        Decision Tree & Maximum depth: 3, 5, 7 \\ \hline
        AdaBoost & Number estimators: 25, 50, 100 \\ \hline
    \end{tabular}
    \caption{Machine Learning algorithms used for classification between experts and novices}
    \label{ml_alg}
\end{table}
The various \gls{ml} algorithm tested are summarized in Table \ref{ml_alg}, with their corresponding parameters ranges. 
Once the best performing \gls{ml} algorithm has been identified, we further optimize it with a grid-search algorithm. Given the small sample size, 10 fold cross-validation (90 percent training/ 10 percent validation set) has been applied with a fixed seed. 

\section{Results}
\begin{figure}[t]
\begin{minipage}[t]{0.48\textwidth}
\begin{tikzpicture}
\begin{axis}[
width=\linewidth,
ybar,
bar width=0.7,
xtick={1,...,7},
xticklabels={2,5,20,100,150,200,400},
tick pos=left,
xlabel={$k_{max}$},
ylabel={Avg. HFD values}
]

\addplot+[error bars/.cd,
y dir=both,y explicit]
coordinates {
    (1,1.02070581077974) +- (0.0, 0.005203282796879)
    (2,1.06566912896654) +- (0.0, 0.01705127259103)
    (3,1.42469115621008) +- (0.0, 0.088894873368078)
    (4,1.67563169301964) +- (0.0, 0.097690440044179)
    (5,1.71235038732567) +- (0.0, 0.090544098879994)
    (6,1.73667075611772) +- (0.0, 0.081516057547179)
    (7,1.80122419071572) +- (0.0, 0.061510680991202)
    };
\end{axis}
\end{tikzpicture}
\caption{HFD value, averaged across all channels all subjects and presentations for different values of $k_{max}$.} \label{average_kmax}
\end{minipage}
\hfill
\begin{minipage}[t]{0.48\textwidth}
\begin{tikzpicture}
\begin{axis}[
width=\linewidth,
ybar,
ymin=0,
bar width=0.7,
xtick={1,...,7},
xticklabels={2,5,20,100,150,200,400},
tick pos=left,
xlabel={$k_{max}$},
ylabel={HFD$_{max}$-HFD$_{min}$}
]

\addplot+[error bars/.cd,
y dir=both,y explicit]
coordinates {
    (1,0.01313700059973) +- (0.0, 0.00397950240834)
    (2,0.041814685837154) +- (0.0, 0.012816160365339)
    (3,0.341177623009121) +- (0.0, 0.076747456159728)
    (4,0.34310607404281) +- (0.0, 0.120405551455622)
    (5,0.337858391110479) +- (0.0, 0.115740850567533)
    (6,0.331669436235316) +- (0.0, 0.108050520895692)
    (7,0.297139917108618) +- (0.0, 0.094485311013287)
    };
\end{axis}
\end{tikzpicture}
\caption{HFD value, difference between HFD values between the maximum and minimum of all channels averaged across all subject and presentations for different values of $k_{max}$.}
\label{maxmin_kmax}
\end{minipage}
\end{figure}


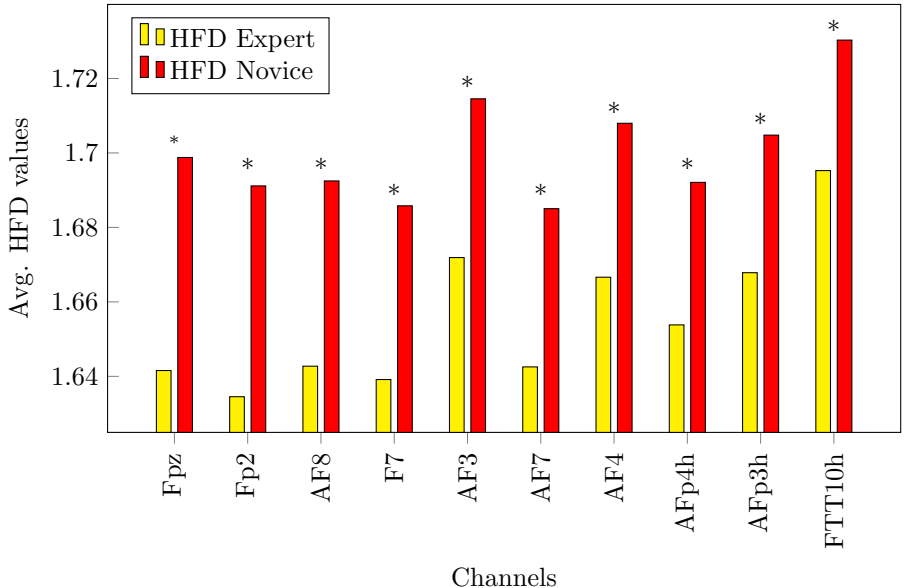
\begin{figure}
    \centering
    \begin{tikzpicture}
    \begin{axis}[
    ybar,
    width=\linewidth,
    height=0.6\linewidth,
    tick pos=left,
    xtick={1,...,10},
    xticklabels={Fpz, Fp2, AF8, F7, AF3, AF7, AF4, AFp4h, AFp3h, FTT10h},
    xticklabel style={rotate=90},
    xlabel={Channels},
    ylabel={Avg. HFD values},
    bar width=0.2,
    legend cell align=left, 
    legend style={legend pos=north west}
    ]
    \addplot+[fill=yellow,draw=black] coordinates {
    (0.99,1.64156681292591)
    (1.99,1.63453047456982)
    (2.99,1.64273165479542) 
    (3.99,1.63914791864488)
    (4.99,1.67190349475678)
    (5.99,1.6425394833014)
    (6.99,1.66662830369965)
    (7.99,1.65381832830175)
    (8.99,1.66784224742322)
    (9.99,1.69526015330037)
    };
    \addplot+[fill=red,draw=black] coordinates {
    (1.01,1.69878290474645)
    (2.01,1.69115129876647)
    (3.01,1.69248424909863) 
    (4.01,1.68580488421934)
    (5.01,1.71453620275145)
    (6.01,1.6850323793294)
    (7.01,1.70796789080717)
    (8.01,1.69209956975593)
    (9.01,1.7047890733084)
    (10.01,1.73030782656063)
    };
    
    \node [above, black, font=\scriptsize] at (axis cs: 1,1.7) {$\ast$};
    \node [above, black, font=\normalsize] at (axis cs: 2,1.693) {$\ast$};
    \node [above, black, font=\normalsize] at (axis cs: 3,1.693) {$\ast$};
    \node [above, black, font=\normalsize] at (axis cs: 4,1.686) {$\ast$};
    \node [above, black, font=\normalsize] at (axis cs: 5,1.715) {$\ast$};
    \node [above, black, font=\normalsize] at (axis cs: 6,1.686) {$\ast$};
    \node [above, black, font=\normalsize] at (axis cs: 7,1.708) {$\ast$};
    \node [above, black, font=\normalsize] at (axis cs: 8,1.693) {$\ast$};
    \node [above, black, font=\normalsize] at (axis cs: 9,1.705) {$\ast$};
    \node [above, black, font=\normalsize] at (axis cs: 10,1.73) {$\ast$};
    
    \legend{HFD Expert,HFD Novice}
    \end{axis}
    \end{tikzpicture}
    \caption{Top 10 channels with the highest difference between their HFD values. Asterisk $\ast$ indicates that the average value of HFD expert is statistically different than HFD novice under p=0.05 threshold for that specific channel.}
    \label{fig:hfda_channels}
\end{figure}
As described in the introduction, extracting the neural signature of math experts and novices requires careful features extraction via the HFD method. To calculate the HFD correctly, hyperparameter $k_{max}$ requires finetuning. Therefore, section \ref{optimalkmax} presents the optimization results of hyperparameter $k_{max}$. Based on the extracted HFD features, experts and novices are compared in section \ref{HFDfeatureselection} giving insights which brain region is relevant for performing mathematical tasks. Finally, based on the features, classification results between experts and novices are shown in section \ref{classification}.

\subsection{Optimal $k_{max}$}  \label{optimalkmax}
Figure \ref{average_kmax} shows the value of HFD for all subjects averaging over all channels for different values of $k_{max}$. HFD is steadily increasing but starts to plateau at a value of 100.
Figure \ref{maxmin_kmax} shows the difference between the maximum and minimum HFD values for different $k_{max}$ with accordance to equation \ref{equ:diff}. It can be observed that the difference in HFD value corresponding to $k_{max}$ reaches a peak at 20 and 100 and progressively declines with increasing $k_{max}$. 
Based on the fact that HFD is plateauing at $k_{max}$ equal to 100 and the largest difference between the maximum and minimum HFD values is also found at the same value, $k_{max}=100$ is used for further analysis.


\subsection{HFD feature analyses}  \label{HFDfeatureselection}
\begin{figure}[!h]
\begin{center}
   \includegraphics[width=0.9\linewidth]{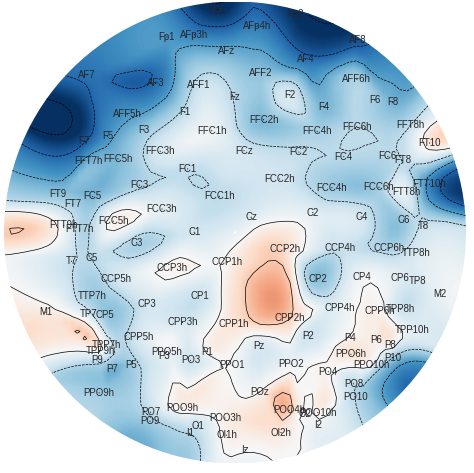}
\end{center}
   \caption{Heatmap of HFD difference between Expert and Novices. Darker blue color indicates the brain areas where the positive differences between experts and novices are the highest.}
\label{topomap}
\end{figure}

Figure \ref{fig:hfda_channels} shows the difference between the average HFD values between experts and novices, for the top 10 channels that present the highest difference between expert and novices. All top 10 channels are statistically significant under p = 0.05 constraint. All channels are depicted in form of a heatmap in Figure \ref{topomap}. The dark blue shaded areas indicate the highest positive difference between expert and novices.

The subsequent more finegrained analysis comparing the difference between expert and novice for algebraic and geometric is shown in Figure \ref{fig:ag_plot} given equation \ref{equ:5}. Although there are differences between algebraic and geometric presentations, none of them is statistically different under p-value 0.05 hypothesis.

\begin{landscape}
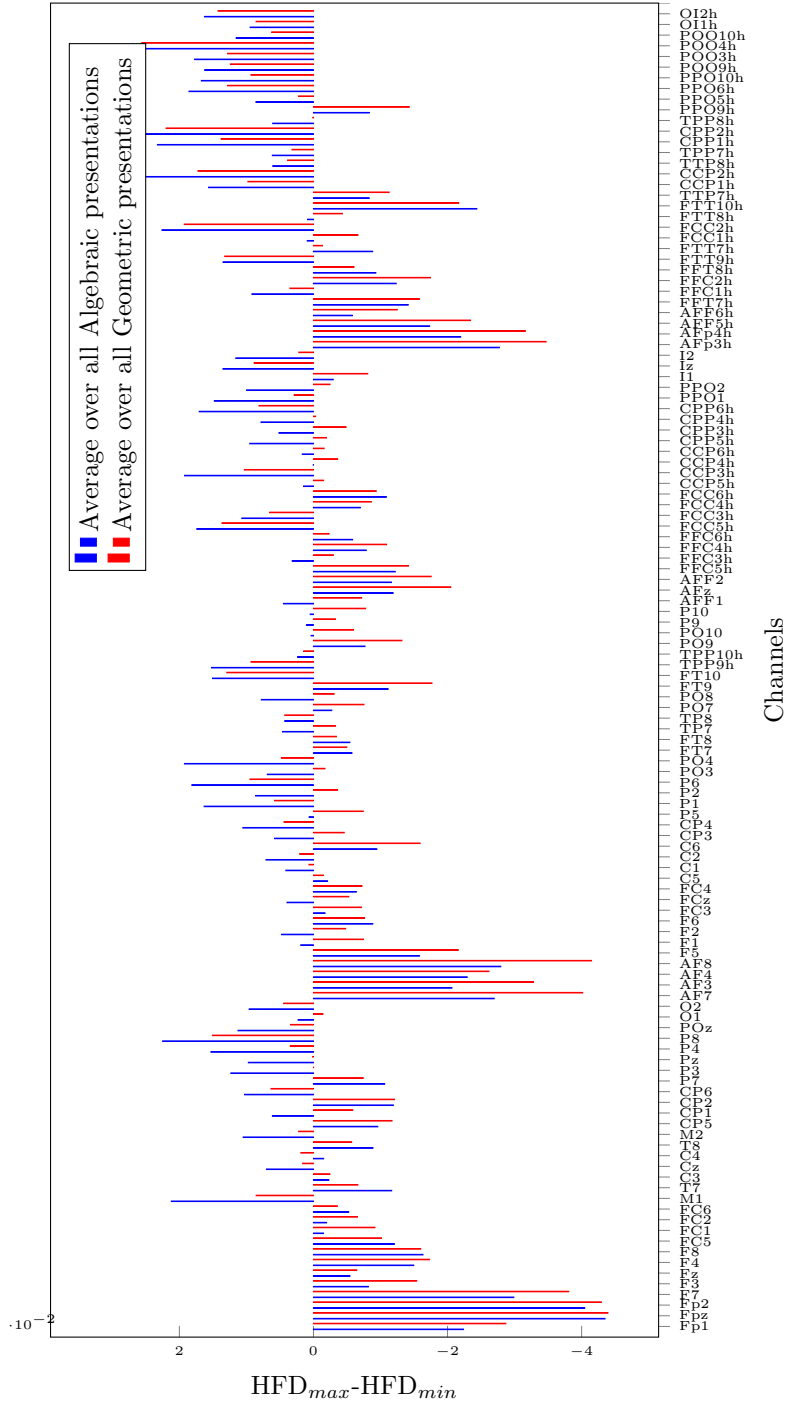
\begin{figure}[!h]
    \centering
    \begin{tikzpicture}[]
    \begin{axis}[
    width=\linewidth,
    height=0.5\linewidth,
    ybar,
    bar width=0.05,
    ticklabel style = {font=\tiny},
    xticklabel style={rotate=-90},
    xtick={0,...,127},
    ylabel={HFD$_{max}$-HFD$_{min}$},
    ylabel style={rotate=180},
    yticklabel style={rotate=-90},
    xlabel={Channels},
    enlarge x limits = {abs = 1},
    xticklabels={Fp1,Fpz,Fp2,F7,F3,Fz,F4,F8,FC5,FC1,FC2,FC6,M1,T7,C3,Cz,C4,T8,M2,CP5,CP1,CP2,CP6,P7,P3,Pz,P4,P8,POz,O1,O2,AF7,AF3,AF4,AF8,F5,F1,F2,F6,FC3,FCz,FC4,C5,C1,C2,C6,CP3,CP4,P5,P1,P2,P6,PO3,PO4,FT7,FT8,TP7,TP8,PO7,PO8,FT9,FT10,TPP9h,TPP10h,PO9,PO10,P9,P10,AFF1,AFz,AFF2,FFC5h,FFC3h,FFC4h,FFC6h,FCC5h,FCC3h,FCC4h,FCC6h,CCP5h,CCP3h,CCP4h,CCP6h,CPP5h,CPP3h,CPP4h,CPP6h,PPO1,PPO2,I1,Iz,I2,AFp3h,AFp4h,AFF5h,AFF6h,FFT7h,FFC1h,FFC2h,FFT8h,FTT9h,FTT7h,FCC1h,FCC2h,FTT8h,FTT10h,TTP7h,CCP1h,CCP2h,TTP8h,TPP7h,CPP1h,CPP2h,TPP8h,PPO9h,PPO5h,PPO6h,PPO10h,POO9h,POO3h,POO4h,POO10h,OI1h,OI2h},
    tick pos=left,
    legend cell align=left, 
    legend style={legend pos=north east}
    ]
    \addplot+[fill=blue,draw=blue] coordinates {
(0,-0.0223838121319159)
(1,-0.0434885534347684)
(2,-0.0404579455806077)
(3,-0.0298802180228406)
(4,-0.00818304537376049)
(5,-0.00544442433616957)
(6,-0.0149573306209924)
(7,-0.0163373934772061)
(8,-0.0120804652981763)
(9,-0.00150002515376985)
(10,-0.0019573983040046)
(11,-0.00525652282819819)
(12,0.0211949413085906)
(13,-0.0116700445042904)
(14,-0.00229344768812026)
(15,0.00700579678058086)
(16,-0.00151296216790747)
(17,-0.00887640898135908)
(18,0.0104803617170441)
(19,-0.00959196837359325)
(20,0.00612088218276941)
(21,-0.0119096613096333)
(22,0.0102692605970489)
(23,-0.0106300200093655)
(24,0.0123185810102221)
(25,0.00971016146481329)
(26,0.0152918575940394)
(27,0.0224863333958812)
(28,0.0112410332971662)
(29,0.00229284761657353)
(30,0.00958917783018884)
(31,-0.0269555405741049)
(32,-0.0206477180729123)
(33,-0.022926577420623)
(34,-0.0279351593552751)
(35,-0.0158041913768359)
(36,0.00189347483238098)
(37,0.00477151502579176)
(38,-0.00885063495770994)
(39,-0.0017241429430245)
(40,0.00393677377084725)
(41,-0.00638724101700716)
(42,-0.00210434475566876)
(43,0.00411587327909016)
(44,0.00707699712675161)
(45,-0.009450712033199)
(46,0.00578617888651947)
(47,0.0105286985293585)
(48,0.000644978856196854)
(49,0.0163046924070562)
(50,0.00861826000072544)
(51,0.0181397406567487)
(52,0.00688180375015801)
(53,0.0192426767662173)
(54,-0.00574666765842127)
(55,-0.00542372810069328)
(56,0.00461561235368932)
(57,0.00428587483348839)
(58,-0.00272766085880596)
(59,0.00777121219421362)
(60,-0.0111409980072909)
(61,0.0150379962290175)
(62,0.0151985869864247)
(63,0.00237292167970893)
(64,-0.00768759370238331)
(65,0.000354548426535262)
(66,0.0010446571708731)
(67,0.000481902801360679)
(68,0.00446979660349883)
(69,-0.011887312287895)
(70,-0.0116202994496161)
(71,-0.0121920288148584)
(72,0.00317461450222734)
(73,-0.00787565846417526)
(74,-0.00583913175123255)
(75,0.0174010762456819)
(76,0.0106810664245544)
(77,-0.00698864246693856)
(78,-0.0108733288056981)
(79,0.00146915001090231)
(80,0.019230678775317)
(81,0.0000248677223498028)
(82,0.00166656654040295)
(83,0.00949596678561701)
(84,0.00514954489215522)
(85,0.00782399236750103)
(86,0.0170380217507294)
(87,0.0147837763918321)
(88,0.00995957800572672)
(89,-0.00294289789123245)
(90,0.0134965513742409)
(91,0.0115805984419114)
(92,-0.0277226803474927)
(93,-0.0219592475292598)
(94,-0.0173281475954541)
(95,-0.00579974067070685)
(96,-0.0141301131177512)
(97,0.00917276386579979)
(98,-0.0123338470788583)
(99,-0.00930283444649885)
(100,0.0134775836767109)
(101,-0.00882957541096246)
(102,0.000925637711088473)
(103,0.0225914801635844)
(104,0.000865634700212309)
(105,-0.0243678884109002)
(106,-0.00831961018346336)
(107,0.0156317221169145)
(108,0.0259964028541646)
(109,0.00606885489865794)
(110,0.00613028518687603)
(111,0.0232696261369231)
(112,0.0307891006650645)
(113,0.006072178543608)
(114,-0.00833905901639667)
(115,0.00858624313126946)
(116,0.018566358799516)
(117,0.0167205336975849)
(118,0.0162268335006923)
(119,0.017714615285588)
(120,0.0316559948742317)
(121,0.0115313529838354)
(122,0.00944168588579747)
(123,0.0162554578639581)
    };
    \addplot+[fill=red,draw=red] coordinates {
(0,-0.0286946421037081)
(1,-0.0439233419263177)
(2,-0.042937080067719)
(3,-0.0380796774643876)
(4,-0.0153996567311115)
(5,-0.00645984804600658)
(6,-0.0173248477127139)
(7,-0.0160004592200121)
(8,-0.0101330372342116)
(9,-0.00917437045456673)
(10,-0.0065631893111289)
(11,-0.00356419391749685)
(12,0.00854507166278787)
(13,-0.0066217006386495)
(14,-0.00244510200400985)
(15,0.00162270240966506)
(16,0.00187577691919813)
(17,-0.00570003875351538)
(18,0.002204417792324)
(19,-0.0117160576178553)
(20,-0.00586294964586709)
(21,-0.0120747237319526)
(22,0.00633198907532179)
(23,-0.00742394418682391)
(24,-0.0000242843623794453)
(25,0.000144031581228138)
(26,0.00346365173714899)
(27,0.0150704810556476)
(28,0.00342674241870072)
(29,-0.00141175461449647)
(30,0.00443721826043517)
(31,-0.0401519929703214)
(32,-0.0328250914090902)
(33,-0.0261739819821045)
(34,-0.0414738073664652)
(35,-0.0216052703548893)
(36,-0.00744378664744735)
(37,-0.00480214677072607)
(38,-0.00762732732092153)
(39,-0.00715633116979275)
(40,-0.00527041224710814)
(41,-0.00723938913398156)
(42,-0.0014734960960423)
(43,0.000666510958423461)
(44,0.00204919528656167)
(45,-0.0159027193665053)
(46,-0.0045852339385403)
(47,0.00439019365683269)
(48,-0.00742104152575693)
(49,0.00580103964856898)
(50,-0.00360666035212306)
(51,0.00947000901504677)
(52,-0.00172193067622364)
(53,0.00478606987302038)
(54,-0.00496928446000336)
(55,-0.00342566455717017)
(56,-0.00329385620636211)
(57,0.00427332898595298)
(58,-0.0075312229281701)
(59,-0.00308645915214398)
(60,-0.0176739964239043)
(61,0.0129200348881159)
(62,0.00932254116669215)
(63,0.00146940635241113)
(64,-0.0131928025185053)
(65,-0.00598674652026732)
(66,-0.00328545218525317)
(67,-0.0077889648077458)
(68,-0.00718950119796011)
(69,-0.0204836244137268)
(70,-0.0175509666669738)
(71,-0.0141772510440775)
(72,-0.00297239621419346)
(73,-0.010902727321791)
(74,-0.00233341724111005)
(75,0.013642386909842)
(76,0.00653231238067514)
(77,-0.00862592496992273)
(78,-0.00936260057145449)
(79,-0.00151404984186176)
(80,0.0103159955614567)
(81,-0.00361078943822848)
(82,-0.00158685439912443)
(83,-0.00194126049804766)
(84,-0.00486784073188778)
(85,-0.000322238763218935)
(86,0.00812774269373107)
(87,0.0028823500412391)
(88,-0.00247009933007467)
(89,-0.00809636180349999)
(90,0.00885127269770208)
(91,0.00217999304691088)
(92,-0.0347009661643448)
(93,-0.0316165303245226)
(94,-0.0234242877896787)
(95,-0.0124966558414389)
(96,-0.0158124311330337)
(97,0.00349954864953894)
(98,-0.0174327478361082)
(99,-0.00605522465124445)
(100,0.0132308004873586)
(101,-0.00132798722325567)
(102,-0.00663783281823699)
(103,0.0192681415711411)
(104,-0.00430187210037503)
(105,-0.02167355516102)
(106,-0.011283849346642)
(107,0.00977154597944546)
(108,0.0172321446404255)
(109,0.00385436128894223)
(110,0.00319402650638256)
(111,0.0137800992073818)
(112,0.0219748036936204)
(113,0.000099581509342056)
(114,-0.0142938838906351)
(115,0.00224090617471881)
(116,0.0127951571164168)
(117,0.00932622162277458)
(118,0.0123925169315404)
(119,0.0128121178396156)
(120,0.0256011337877662)
(121,0.00622868517106495)
(122,0.0085534194024737)
(123,0.0142413227297558)

    };
    \legend{Average over all Algebraic presentations, Average over all Geometric presentations}
    \end{axis}
    \end{tikzpicture}
    \caption{HFD$_{max}$-HFD$_{min}$ calculated as average for all Algebraic vs Geometric presentations for all channels as defined in equation \ref{equ:5}}
    \label{fig:ag_plot}
\end{figure}
\end{landscape}

\subsection{Expert/Novice classification} \label{classification}

Table \ref{tab:ml_results} summarizes the classification results between expert and novices. On a subject-presentation split the accuracy reaches 97\% demonstrating that it is possible to automatically classify between math experts and math novices based on their \gls{eeg} signals while watching math demonstrations because the \gls{ml} model can successfully learn each subject's brainwaves signatures. 

\begin{table}[h!]
\centering
\resizebox{\columnwidth}{!}{%
\begin{tabular}{l|l|l|llllllllllllll}
                  & Subject- & Subject & \multicolumn{14}{c}{Presentation specific split}                                            \\
                  & pres. pairs        & specific   & 1A   & 1G   & 2A   & 2G   & 3A   & 3G   & 4A   & 4G   & 5A   & 5G   & 7A   & 7G   & 9A   & 9G   \\ \hline
Nearest Neighbors & 97\%         & 48\%    & 48\% & 61\% & 52\% & 51\% & 49\% & 45\% & 60\% & 47\% & 53\% & 66\% & 67\% & 57\% & 43\% & 55\% \\
Linear SVM        & 87\%         & 66\%    & 56\% & 53\% & 53\% & 54\% & 46\% & 61\% & 49\% & 43\% & 63\% & 46\% & 63\% & 49\% & 39\% & 61\% \\
Decision Tree     & 78\%         & 50\%    & 49\% & 47\% & 55\% & 59\% & 52\% & 53\% & 51\% & 41\% & 55\% & 50\% & 79\% & 49\% & 47\% & 42\% \\
AdaBoost          & 93\%         & 56\%    & 53\% & 51\% & 45\% & 51\% & 49\% & 59\% & 52\% & 39\% & 66\% & 59\% & 61\% & 39\% & 47\% & 56\% \\ \hline
\textbf{Best} &
  \textbf{97\%} &
  \textbf{66\%} &
  \textbf{56\%} &
  \textbf{61\%} &
  \textbf{55\%} &
  \textbf{59\%} &
  \textbf{52\%} &
  \textbf{61\%} &
  \textbf{60\%} &
  \textbf{47\%} &
  \textbf{66\%} &
  \textbf{66\%} &
  \textbf{79\%} &
  \textbf{57\%} &
  \textbf{47\%} &
  \textbf{61\%}
\end{tabular}%
}
\caption{Classification results between experts and novices based on different classification algorithms for Subject-presentation pairs, Subject-specific and Presentation-specific split. All results are based 10 fold cross validation and averaged over 3 random seeds.}
\label{tab:ml_results}
\end{table}

However, when we split the training and test sets on a subject level, meaning that we increase the difficulty of the task by introducing inter-subject variability that is well-known to be challenging in biosignals classification, i.e., the trained model is validated on new subjects whose data it has never seen before, the accuracy falls to 66\%. 

So far the results are shown by considering all presentations for each subject, i.e., the calculated HFD features for all presentations are concatenated for the final classification stage. We suspect that the poor classification accuracy could be partially caused by some of the presentations that might perform poorly.
Hence, we perform presentation-specific classification on subject level and the classification accuracy improves up to 79\% (presentation 7A).

\begin{figure}[t]
\begin{minipage}[t]{0.51\textwidth}
\includegraphics[width=\linewidth]{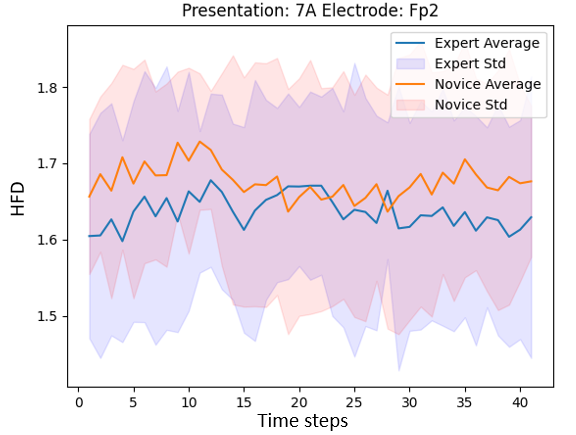}
\caption{HFD values (before averaging) for presentation 7A, channel FP2 for Experts (average) and Novices (average)} \label{channel7A}
\end{minipage}
\hfill
\begin{minipage}[t]{0.49\textwidth}
\includegraphics[width=\linewidth]{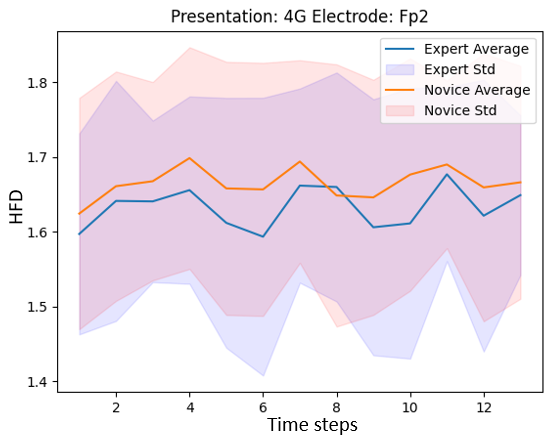}
\caption{HFD values (before averaging) for presentation 4G, channel FP2 for Experts (average) and Novices (average)}\label{channel4G}
\end{minipage}
\end{figure}

Figure \ref{channel7A} and Figure \ref{channel4G} show the HFD values when window size of 8 seconds is applied for the presentation with the highest (presentation 7A) and the lowest (presentation 4G) classification accuracy. The difference in classification accuracy may be explained through a better separation between Experts and Novices.

\subsection{Discussion}

Advantages of ML for brain research include the data driven approach which enables generation of hypotheses about underlying brain processes in rest or in active engagement with a cognitive or emotional task. Such underlying processes are sometimes impossible to detect by experts’ observations. ML also enables explorations of new paradigms with respect to their neurophysiological signatures (Lemm et al., 2011). One of such new paradigms is naturalistic study design which aims to understand the brain during real-life tasks, like when solving complex math. 

Our novel approach on applying ML to EEG data recorded in math experts and novices during complex math encourages to expand the usage of data driven brain imaging methods from healthcare to education. Our approach utilizing nonlinear HFD, which measures signal complexity, was reliable in describing the data by systematically detecting the difference in the neural signature of math experts and novices with a 98\% cross-validation accuracy. However, the results gained with ML discriminative algorithm were mixed and showed 50-80 percent classification accuracy when tested with unseen subjects.

Nonlinear fractal dimension methods seem ideal for tracing fluctuations in biological systems, including the brain, which are nonlinear by nature. HFD is a measure of signal complexity in the time domain (Higuchi, 1988 \cite{HIGUCHI1988277}; Spasic et al., 2008 \cite{Spasic2008}) and has been successfully applied for brain state analysis of EEG in sleep, drowsiness, and wakefulness (Inouye et al., 1994 \cite{Inouye1994ChangesIT}; Klonowski et al., 2005 \cite{Klonowski2002}; Peiris et al., 2006). Our results gained with HFD show a difference in the neural signature between math experts and novices during long and complex math tasks with a high classification accuracy. These results encourage to use the HFD method in detecting subtle differences in the brain states, like those of math experts and novices, which go beyond the more drastic differences in the brain states during the levels of arousal, like sleep stages, or drowsiness and wakefulness.    

Despite the successful classification to experts and novices based on HFD was relatively stable for the entire dataset, the ML model adapted poorly to unseen subjects, and we could not overcome the overfitting and high generalization error caused by inter-subject variability. The most important reason for such a poor generalization is that our dataset is incorrigibly small to be divided to the training and test sets on a subject level. 
In healthcare, big data platforms are being formed increasingly (Eickhoff et al., 2016; Zbontar et al., 2019), and it is important to take similar steps to create large and clearly labeled open data pools for educational neurosciences.  

Our small dataset may function reasonably well for method development of data-driven approaches, since the differences between math demonstrations are statistically significant especially over several frontal electrodes showing higher frontal signal complexity in math novices in comparison to experts. Cognitively, these results may indicate novices’ stronger recruitment of domain-general processes in comparison to experts, which is in line with previous literature (Amalric and Dehaene, 2016 \cite{Amalric2016}; Wang et al., 2020 \cite{Wang2020}). 

Some studies have investigated for the connection between nonlinear FD methods and linear oscillation analyses over delta, theta and alpha bands. These studies show a dependence between the nonlinear and linear methods and suggest that the most reliable results are gained when combining nonlinear and linear methods to classify different brain states (Acharya et al., 2005 \cite{Amalric2016}; Šušmáková and Krakovská, 2008 \cite{Susmakova2008}). Since combination of nonlinear and linear methods seem to bring the most robust classification results, we could combine the HFD and oscillation analyses and feed the combined information to a machine learning model. Our novel analysis with machine learning utilized only fractal dimension; however, we report on other papers the brain oscillations for the same dataset (Formaz et al., unpublished data; Poikonen et al., 2022 \cite{poikonnen2022}).  

Another interesting way to deepen the analysis of our dataset was to break the temporal data stream to segments. With a larger dataset and statistical power, time points during which the neural signatures of math experts and novices differ the most could potentially be found. This data-driven approach may have practical implications after detecting whether the cortical functions of experts and novices differ the most at the beginning, at the end, or at some other time point during the math demonstrations. With our dataset, ML algorithm showed 50-80 percent classification accuracy for unseen subjects when breaking the data to a temporal stream. Such a high variation may be explained by a small dataset, or by a combination several features related to the length, content, and difficulty level of the math demonstrations.  

Understanding which parts of the math demonstrations to emphasize when teaching complex math may be helpful in supporting students’ development towards math expertise. Such time-dependent information may be hard to collect with questionnaires or other behavioral measures, and therefore, brain-originated data-driven methods may be the only way to access such information in the context of learning. Further, these ML models could be used to create learning contexts in which adaptive feedback is given to adjust to the individual needs of a learner or those of a specific group during collaborative learning, building on the previous examples like BCI applications for post-stroke motor rehabilitation, or relatively simple neurofeedback applications for focused attention or working memory (Cervera et al., 2018 \cite{Cervera2017BrainComputerIF}; Kefalis et al., 2020; Hunkin et al., 2021 \cite{Hunkin2020EEGND}). Simple options for BCI interventions for the math demonstrations used in our study might be to adjust the velocity of presenting new information, or by scaffolding the learning process via instructions or remarks depending on the EEG signal of the learner.

\section{Limitations} 

Our novel paradigm combining mathematical cognition, cortical activity and ML is exploratory in nature and we recognize the following limitations. First, the most drastic limitation is the small dataset in use. The straightforward way around it would be to increase significantly the amount of data, e.g., by at least doubling the number of participants. The more data the better we can estimate the real data distribution of the general population. 
The second limitation is related to the classes chosen for the ML classification. We chose to compare two groups of participants during the same cognitive task. Other strategy for a small dataset would be to explore individual differences, for example, by aiming to classify the data excerpts of resting state and cognitively active state for each participant. Earlier studies show that differentiation of brain states for an individual participant during simple sensory tasks is rather robust whereas the generalizations of the cortical activation patterns across a group of participants, and during complex cognitive tasks, is challenging. However, such individual brain state classification would not give us hardly any insights for the expert-novice differences during mathematical cognition. 
As the third limitation to consider, when preprocessing, we chose to band-pass filter the data with a bandwidth of 0.5-40 Hz due to the contamination of the data with the 50 Hz line noise. HFD is associated with changes in delta, theta and alpha oscillations which all were included in our analysis. However, also gamma oscillation is known to be important during cognitive tasks (Herrmann et al., 2004), and it has been connected to HFD. Due to bandpass filtering chosen, gamma activity is not included in our analysis. 
Based on previous literature, HFD seems the most stable fractal dimension methods (Kesic and Spasic, 2016 \cite{KESIC201655}). However, as the fourth limitation of our study, is the general criticism for the HFD that it has a short margin of scale which may give the same complexity number to signals with only subtle differences. For detecting the possibly small differences in the cortical activity of math experts and novices, some other method with more detailed scale may be more suitable. 
Fifth, for the cross-validation, different models could be compared to find a model with ideal complexity which balances between overfitting of an unnecessarily complex model and simple model’s inability to adapt to the details of the complex cognitive data. Ideally for ML algorithms, each sample (e.g. EEG data collected during each math demonstration) would have the same number of data points (e.g. the same duration). However, it is difficult to realize in practice due to different duration it takes to solve different naturalistic math tasks. In the future, research of brain processes during abstract cognition might be conducted, for example, within a video game context, in which the duration is easier to match to be the same over all the rounds played.    

Conclusions
The present study used a unique paradigm to compare neural correlates of math experts and novices while solving naturalistic math demonstrations. Overcoming limitations of previous studies with reductionist stimuli and linear EEG analysis methods, the brain functions during abstract cognition were measured with a high-density EEG during long and complex math demonstrations and analyzed with a relatively rigor nonlinear method, HFD. Our results indicated that math novices have a higher signal complexity measure with HFD than experts over several frontal electrodes suggesting a stronger engagement of domain-general brain functions. Further, we explored ML algorithms for classifying math experts and novices based on their neural signature. These results were promising but we also acknowledge the inevitably small dataset we had in use for consistent results. We encourage taking example from brain imaging databases created in healthcare for a creation of a similar database for educational neuroscience. In the future, application possibilities for such a database and deep learning lay in data-driven theory formation for normal and disrupted learning and development, and adaptive feedback systems for learning contexts.

Data and code availability statement
The dataset analyzed in the present study as well as scripting and plotting code are available from the corresponding authors via email on request.

CRediT (Contributor Roles Taxonomy) authorship contribution statement (see https://www.elsevier.com/authors/policies-and-guidelines/credit-author-statement)
Hanna Poikonen: Conceptualization, Methodology, Investigation, Funding acquisition, Data Curation, Writing – original draft, Writing – review \& editing. Tomasz Zaluska: Formal analysis, Visualization, Writing – original draft, Writing – review \& editing. Xiaying Wang: Methodology, Supervision, Writing – review \& editing. Michele Magno: Supervision, Methodology, Writing – review \& editing. Manu Kapur: Conceptualization, Supervision, Writing – review \& editing.

Funding
This work was supported by a grant from the Ella and Georg Ehrnrooth Foundation awarded to H.P.

Declaration of Competing Interest
The authors have no conflict of interest to disclose.

Acknowledgments
We thank Dr. Dragan Trninic and Dr. Venera Gashaj for collaboration with the creation of the math stimuli. We also thank Cléa Formaz, Samuel Tobler, Maya Spannagel and Lea Imhof for their help in data acquisition and Stefan Wehrli for all the help with the NeuroLab of the Decisions Sciences lab.

\bibliography{mybibfile}

\end{document}